\documentclass[10pt,conference]{IEEEtran}

\usepackage{graphicx}
\usepackage{amssymb}
\usepackage{amsmath}
\usepackage{cite}
\usepackage{multirow}
\usepackage{color}

\begin{document}
\title{Integer Linear Programming Modeling of Addition Sequences With Additional Constraints for Evaluation of Power Terms}

\author{\IEEEauthorblockN{Muhammad Abbas and Oscar Gustafsson}
\IEEEauthorblockA{Department of Electrical Engineering, Link\"oping University\\
SE-581 83 Link\"oping, Sweden\\
E-mail: \texttt{oscar.gustafsson@liu.se}}}
\maketitle

\begin{abstract}
In this work, an integer linear programming (ILP) based model is proposed for the computation of a minimal cost addition sequence for a given set of integers. Since exponents are additive under multiplication, the minimal length addition sequence will provide an economical solution for the evaluation of a requested set of power terms. This is turn, finds application in, e.g., window-based exponentiation for cryptography and polynomial evaluation.  Not only is an optimal model proposed, the model is extended to consider different costs for multipliers and squarers as well as controlling the depth of the resulting addition sequence.
\end{abstract}

\IEEEpeerreviewmaketitle

\section{Introduction}

Exponentiation is a fundamental arithmetic operation that finds applications in, e.g., computational number theory \cite{Cruz-Cortes2008,Gordon1998}, cryptography \cite{Rivest1978, Tillborg2005}, and polynomial evaluation \cite{Dobkin1980, Dinechin2010}. For a positive integer $n$, exponentiation can be realized as repeated multiplications. The most straightforward way to compute $x^n$ is to do $n-1$ multiplications. However, for large values of $n$, this would be infeasible to compute $x^n$ using $n-1$ successive multiplications by $x$. However, it is easy to see that using more than one intermediate result in each iteration may reduce the number of operations. Consider the computation of $x^4$ which can be realize as $(x\times x)\times(x\times x)$ rather than $x\times(x \times (x \times x))$, reducing the number of multiplications from three to two, as $x \times x$ only needs to be computed once. In addition, these two multiplications are in fact squarers, which have a significantly lower complexity compared to general multiplications \cite{Pihl1996}. 

To effectively evaluate a power $x^n$, it would be of interest to find an algorithm using as few multiplications as possible. This problem is often referred to as the addition chain problem, as exponents are additive under multiplication, the power evaluation problem is equivalent of finding an ascending list of integers such that any element of this list except the first one can be represented as the sum of two preceding elements of the list \cite{Knuth1998}.

In this work, we are mainly interested in computing a set of integer powers using as few multiplications as possible. This is a generalization of the addition chain problem, called the addition sequence problem \cite{Olivos1981}. The addition sequence problem is related to the constant multiplication problem arising in DSP algorithms \cite{Gustafsson2007}.

Computing a set of integer powers is used in windowing-based exponentiation. Instead of finding the minimum addition chain for a long number, such as the keys used in some cryptographic algorithms, a good approximation is to look at a smaller set of bits, a window, and compute an exponent corresponding to the value in that window. These exponents can then be combined using squarers to end up with an efficient exponentiation algorithm. Typically, not all possible exponents for a given window-size are required. A set of powers may also be computed in the case of evaluating a sparse polynomial.

Several different heuristics have been proposed for the addition sequence problem (and even more for the addition chain problem). However, so far, no optimal algorithms have been proposed, which is partly related to the fact that both the addition chain and addition sequence problems are NP-hard. 

In this paper, we propose an integer linear programming (ILP) modeling for finding optimal addition sequences. Furthermore, we discuss modifications of the model to allow control of the number of squarers as well as the number of cascaded operations (the depth). Some useful cuts are also introduced to decrease the solution time.

In the next section, a brief review of addition chains and addition sequences is presented along with their upper and lower bounds.
\section{Addition Chains and Addition Sequences}
An addition chain for an integer number $n$ is an ascending list of integers such that any element of this list except the first one can be represented as the sum of two preceding elements of the list. An addition chain for $n$ is given by a list of positive integers as \cite{Knuth1998, Bleichenbacher1998}
\begin{equation}
v_1=1, \quad v2, \dots, v_s=n, 
\end{equation}
such that, for each value of $i > 1$, there is some $j$ and $k$ with $1\le j \le k <i$ and
\begin{equation}
v_i=v_j+v_k.
\end{equation}
A short addition chain for any positive integer $n$ gives the fast method for computing any power raised to that integer value. The length of an addition chain is $s$. For a given $n$, the smallest $s$ for which there exists an addition chain of length $s$ computing $n$ is denoted by $l(n)$. The determination of $l(n)$ is a difficult problem even for small values of $n$ \cite{Downey1981}. A lower bound on the length of addition chains is \cite{Knuth1998, Schonhage1975}
\begin{align}
l(n) \ge \log_2(n) + \log_2(g(n))-2.13, \notag
\end{align}
and the upper bound is
\begin{align}
l(n) \le \lfloor \log_2(n) \rfloor + \log_2(g(n))-1, \notag
\end{align}
where $g(n)$ is the number of ones in the binary representation of $n$. A better upper bound from \cite{Brauer1939} is
\begin{align}
l(n) \le \log_2(n) + \frac{\log_2(n)}{\log_2(\log_2(n))}+ O\left(\frac{\log_2(n)}{\log_2(\log_2(n))}\right). \notag
\end{align}
A brief summary of prior results and useful bounds are given in \cite{Bahig2011}.
 
The concept of addition chains can be extended in many different ways. One of them is the addition sequence, where several numbers should be included in the addition chain. In the case of an addition chain for a given number, the given number in the chain appears at the end. However, in the case of an addition sequence, the given numbers occur in the sequence. The length of an addition chain or sequence is the number of elements in the chain apart from the initial one.  

An addition sequence for the set of integers $T = \{n_1, n_2, \dots , n_r\}$ is an addition chain $v$ that contains each element of $T$. In other words, for all $k$ there is a $j$ such that $n_k = v_j$. For example, an addition sequence computing $\{3, 7, 11\}$ is $(1, 2,$ $\bf{3},$ $4,$ $\bf{7},$ $9,$ $\bf{11})$. In \cite{Yao1976}, it is shown that the shortest length of an addition sequence computing the set of integers $\{n_1, n_2, \dots, n_{r}\}$ is bounded by
\begin{equation}
 l(n_1, n_2, \dots, n_{r}) \le \log_2(N)+ c \left(\frac{\log_2(N)}{\log_2( \log_2(N))}\right)r,\notag
\end{equation}
where $N = \textrm{max} ({n_k})$, $k=1,2, \dots, r$, and $c$ is a constant given by
\begin{equation}
c \approx 2+\frac{4}{\sqrt{n_{r}}}.\notag
\end{equation}

In the next section, the proposed ILP model is described along with the useful constraints, cuts, and other additional aspects to control the number of squarers and depth.
\section{Proposed ILP Model}
An addition sequence of minimal length for the required set of integers would optimize the number of steps required to compute all these numbers in the set. However, finding this minimal addition sequence is an NP-hard problem and there is a need of efficient algorithms to solve it. Known techniques for the addition chain problem do not combine well in eliminating the overlapping operations in the evaluation of a set of integers \cite{Wojko1999}. 

The basic ILP model is described first with its methodology and basic constraints. The additional aspects and extensions will be described later. 
\subsection {Basic ILP Model}
In the proposed ILP model, the problem of determining the minimal length addition sequence is formulated with the use of binary variables. Let $x_k \in \{0,1\}, k \in K = \{1, 2, ..., n_r\}$, be a variable which is one if and only if the value $k$ is within the addition chain. In addition, let $y_{i, j} \in \{0,1\}$ be a variable that is one if and only if the integer numbers $i$ and $j$ in the addition sequence are used to compute another number of the addition sequence, $k=i+j$. The integers $i$ and $j$ are not necessarily required to be  different and can take any combination from set $K$ such that $2 \le i+j \le n_{r}$. While the order of $i$ and $j$ is not important, we would like to have a non-redundant representation, i.e., avoiding to have both $y_{i,j}$ and $y_{j,i}$ as they correspond to the same thing. Hence, by definition we choose $i \leq j$. Based on this, we can define a set $P = \{i,j \in K: i+j \leq n_r, i \leq j\}$ containing all possible combinations of $i$ and $j$ to consider.

The objective function to be minimized is the addition sequence length as
\begin{equation}
\text{minimize: } l(T) = \sum_{(i,j) \in P} y_{i, j}, \notag
\end{equation}
where $l(T)$ is the minimum addition sequence length for the set of powers $T = \{n_1, n_2, \dots , n_r\}$. Note that it is also possible to minimize
\[
\sum_{k \in K} x_k.
\]
However, as we will see later, the $y_{i,j}$ values include information if multiplications or squarers are used.

The most important constraint in the proposed ILP model that controls the generation of any other new number in the addition sequence with the help of numbers already available in the addition sequence is 
\begin{equation}
\sum_{(i,j) \in P:i+j=k} y_{i, j}=x_k,\   \forall k \in K
\end{equation}
To make sure that numbers $i$ and $j$ are already computed and are available before computing $y_{i,j}$, two constraints are added as
\begin{equation}
y_{i, j} \le x_i,\ \forall (i,j) \in P\notag 
\end{equation}
and
\begin{equation}
y_{i, j} \le x_{j},\ \forall (i,j) \in P. \notag 
\end{equation}

Finally, to make sure that all integer numbers of the set $T$ are computed and are in the minimum length addition sequence:
\begin{equation}
x_k=1, \quad \forall k \in T. \notag 
\end{equation}

The basic ILP model gives the correct solution but the solution time may increase rapidly as the problem complexity is increased. To deal with this, a class of cuts are suggested. These are not required for solving the problem, but can possibly improve the solution time. The proposed cut relies on the fact that to compute a term $k$, at least one term of weight $\geq \left\lceil k/2 \right\rceil$ must be computed. This can then be formulated as
\begin{equation}
\text{Cut:} \sum_{m=\lceil k/2\rceil}^{k-1}x_m \ge 1, \quad \forall k \in K.\notag
\end{equation}
  
As we will see in the result section, this cut will on average reduce the solution time. It will also increase the linear relaxation of the problem and sometimes make the linear relaxation a valid integer solution.

\subsection {Minimizing Weighted Cost}
As mentioned earlier, if squarers are available these are preferred before multipliers, as the squarers have a lower cost. However, the previous model does not take the difference between squarers and multipliers into account.

As also mentioned in the previous section, the information about whether a squarer or a multiplier should be used is within the $y_{i,j}$ variables. If $i=j$ then a squarer can be used, otherwise not. Thus, assuming a relative cost of $C_m$ for multipliers and $C_s$ for squarers, the objective function can be written as
\[ \text{minimize: } C_m N_m + C_s N_s, \]
where $N_m$ and $N_s$ are the number of multiplications and squarers, respectively. These can be expressed as

\begin{align}
N_m & = \sum_{(i,j) \in P: i\neq j} y_{i,j}\\
N_s & = \sum_{(i,j) \in P: i = j} y_{i,j} = \sum_{k \in K} y_{k,k}
\end{align}

This means that it is possible to find a solution with a maximum number of squarers by selecting appropriate values of $C_m$ and $C_s$. In the results section, we will assume $C_m=2$ and $C_s=1$ corresponding to the fact that a squarer has roughly half the number of partial products compared to a multiplier and therefore roughly half the area.

\subsection {Minimizing Depth}
To control the depth, a variable $d_k$ denoting the depth of the operation evaluating $k$ is introduced.
Defining the depth of the input to be zero, the basic depth constraints are
\begin{equation}
d_i \le d_k + (1-y_{i,j})(d_{\textrm{max}}+1)-1, \forall k \in K, (i,j) \in P: i+j=k\notag
\end{equation}
and 
\begin{equation}
d_j \le d_k + (1-y_{i,j})(d_{\textrm{max}}+1)-1, \forall k \in K, (i,j) \in P: i+j=k\notag
\end{equation}
where $d_{\max}$ is a parameter determining the maximum allowed depth.

By adding the following constraints, it is possible to limit the unused depth variables to 0, as they otherwise are unconstrained
\begin{equation}
d_k \le d_{\textrm{max}}x_k,\quad 2 \le k \le n_{r}.\notag
\end{equation}
Combining with the following constraints will help reduce the solution time
\begin{equation}
x_k d_{\textrm{min}} \le d_k,\  \forall k \in K,\notag
\end{equation}
where $d_{\textrm{min}}=\lceil \log_{2}(n_r) \rceil$ is the minimum depth.

\section{Results}

\begin{figure}[tb]
  \centering
  \includegraphics[scale=0.40]{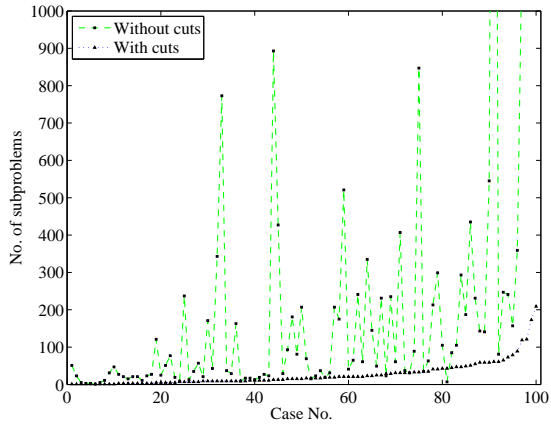}
  \caption{The solution time of the ILP model for different sets of power terms with and without using proposed cuts.}
  \label{fig:time}
\end{figure}

The functionality of the proposed ILP model for minimum length addition sequence is first tested and verified when it is applied to a set of  integer terms, $T=\{23, 41, 67\}$, with already known minimum addition sequence length. All the later simulation of the model are done using a simulation setup in MATLAB/GLPK. First a random number between $1$ and $10$ is generated for the number of power terms in the set excluding the initial one. As per the number of power terms, a set of random power terms with numeric values between $1$ and $63$ is generated to test the ILP model. We have considered $1000$ different sets of power terms with different lengths and numeric values. After going through the verification run, the basic ILP model is tested with and without using cuts for the solution time. The additional cuts are not required for solving the problem but are used to limit the solution time. As can be seen in Fig.~\ref{fig:time}, the solution time is decreased by using additional cuts. The difference, however, will be more significant as the numeric values and the length of the requested set of power terms is increased.

The solution returned from the basic model gives minimum number of operators but since there is no differentiation between the squarers and multipliers, the solution will not be optimized for the cost. The next run is made for the cost minimization by using weighted objection function in order to optimize the solution by maximizing the number of squarers. A weight of two and one is used for multiplier and squarers respectively.  In Table~\ref{tab:cost_sqr}, a list of different sets of power terms is given, where solution with same number of operators is optimized to give same number of operators but with maximum number of squarers. 

\begin{figure}[tb]
  \centering
  \includegraphics[scale=0.45]{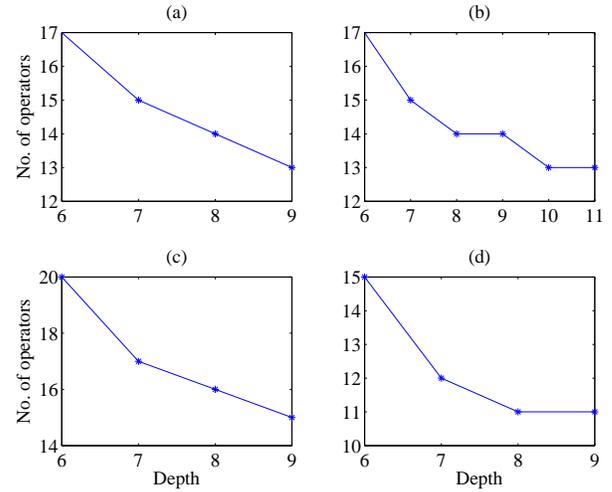}
  \caption{Trade-offs between depth and number of operators for different sets of power terms in Table~\ref{tab:powersDepth}.}
  \label{fig:depth}
\end{figure}

Another interesting and useful aspect added to the basic ILP model is the depth solution. Trade-offs between the number of operators and depth is considered as shown in Fig.~\ref{fig:depth}. Four different cases are studied and power terms for each case are given in Table~\ref{tab:powersDepth}. As expected, for the minimum depth, the number of operators are high. However, when depth constraint is relaxed step-by-step, the operator count is also decreased. The computation of power terms for the set (a) in Table~\ref{tab:powersDepth} at different depths subject to different minimum depth constraints is given in Table~\ref{tab:depth1}. Additional constraints for the depth model to limit their unconstrained values and the processing time are also tested and verified to demonstrate their full potential. 

\begin{table}[ht]
\caption{Optimizing the solution to the maximum number of squarers by minimizing the weighted cost for different sets of power terms. W, WO are for with and without optimizations and M, S are for multipliers and squarers respectively.}
\centering
\begin{tabular}{|c|c|c|c|c|c|}
\hline
 \multirow{2}{*}{Sr. No.} & \multirow{2}{*}{Power terms} & \multicolumn{2}{c|}{WO} & \multicolumn{2}{|c|}{W} \\
\cline{3-6}
       & & M & S & M & S \\
\hline
$1$  & $1, 49, 54, 59$ & $7$ & $3$ & $5$ & $5$ \\
$2$  & $1, 5, 51, 63 $& $6$ & $3$ & $4$ & $5$ \\
$3$  & $1, 55, 59, 63 $& $6$ & $4$ & $5$ & $5$ \\
$4$  & $1, 22, 39, 50 $& $6$ & $3$ & $5$ & $4$ \\
$5$  & $1, 51, 56, 58 $& $7$ & $3$ & $5$ & $5$ \\
$6$  & $1, 27, 50, 58, 61$ & $7$ & $4$ & $5$ & $6$ \\
$7$  & $1, 37, 39, 51, 55, 57 $& $8$ & $3$ & $7$ & $4$ \\
$8$  & $1, 11, 33, 49, 53, 55, 63$ & $8$ & $4$ & $7$ & $5$ \\
$9$  & $1, 37, 44, 52, 56, 59, 63 $& $9$ & $4$ & $7$ & $6$ \\
$10$  & $1, 6, 15, 17, 29, 42, 44, 48 $& $8$ & $3$ & $7$ & $4$ \\
$11$  & $1, 7, 12, 17, 28, 42, 44, 52, 56, 59 $& $10$ & $4$ & $7$ & $7$ \\
$12$  & $1, 6, 12, 15, 17, 24, 36, 47, 59, 61 $& $8$ & $4$ & $7$ & $5$ \\

\hline
 \end{tabular}
\label{tab:cost_sqr}
\end{table}

\begin{table}[t]
  \centering
  \caption{Power terms used in Fig.~\ref{fig:depth} to study the trade-offs between the number of operators and depth.}
  \label{tab:powersDepth}
  \begin{tabular}{l|c}
\hline
Case & Power terms\\
\hline
   (a) & $1, 31, 38, 39, 43, 51, 55, 56$ \\
   (b) & $1, 9, 10, 29, 31, 33, 35, 44, 46, 51$\\
(c) & 1, $10, 22, 27, 37, 39, 40, 46, 48, 53, 61$\\
(d) & $1, 7, 24, 26, 38, 44, 59$\\ 
\hline
 \end{tabular}
\end{table}

\begin{table}[htb]
\caption{Computation of power terms for the set (a) in Table~\ref{tab:powersDepth} at different depths, $D$, subject to step-by-step relaxation of minimum depth constraint from $d_{\textrm{min}}$ to $d_{\textrm{min}}+3$.}
\centering
\begin{tabular}{|c| c| c| c| c|}
\hline
\multirow{2}{*}{$D$} & \multicolumn{4}{|c|}{Power terms at different depths}\\
\cline{2-5}
& $d_{\textrm{min}}$ & $d_{\textrm{min}}+1$ & $d_{\textrm{min}}+2$ & $d_{\textrm{min}}+3$\\
\hline
$0$ & $1$ & $1$ & $1$ & $1$ \\
$1$ & $2$ & $2$ & $2$ & $2$ \\
$2$ & $3, 4$ & $3$ & $4$ & $4$ \\
$3$ & $7, 8$ & $6$ & $6$ & $6$ \\
$4$ & $15, 16$ & $7, 12$ & $7, 12$ & $12$ \\
$5$ & $19, 24, 31, 32$ & $13, 19, 24$ & $19, 24$ & $13$\\
$6$ & $38, 39, 43, 51, 55, 56$ & $31, 38, 43$ & $31, 38, 43$ & $25$ \\
$7$ &  & $39, 51, 55, 56$ & $39, 55$ & $31, 38$ \\
$8$ &  &  &$51, 56$ & $39, 51, 56$ \\
$9$ &  &  & & $43, 55$ \\
\hline
 \end{tabular}
\label{tab:depth1}
\end{table}

\section{Conclusions}
An ILP model of the addition sequence problem was formulated and additional cuts were proposed. Since exponents are additive under multiplication, the minimal length addition sequence would provide an economical solution for the evaluation of a requested set of power terms. The usefulness of cuts were tested and it was found that the solution time with the use of cuts was comparably much lower. The basic ILP model was then extended to consider different costs for multipliers and squarers in order to optimized the solution for maximum number of squarers. Examples were shown where it was possible to optimize the cost by maximizing the number of squarers.  A trade-off between number of operators and depth was found when ILP model was used to control the depth of the resulting addition sequence.

\end{document}